\documentclass{moriond}

\usepackage{amssymb}
\usepackage{comment}
\def\be{\begin{equation}}
\def\ee{\end{equation}}
\def\bea{\begin{eqnarray}}
\def\eea{\end{eqnarray}}

\newcommand{\Photo}{\includegraphics[height=35mm]{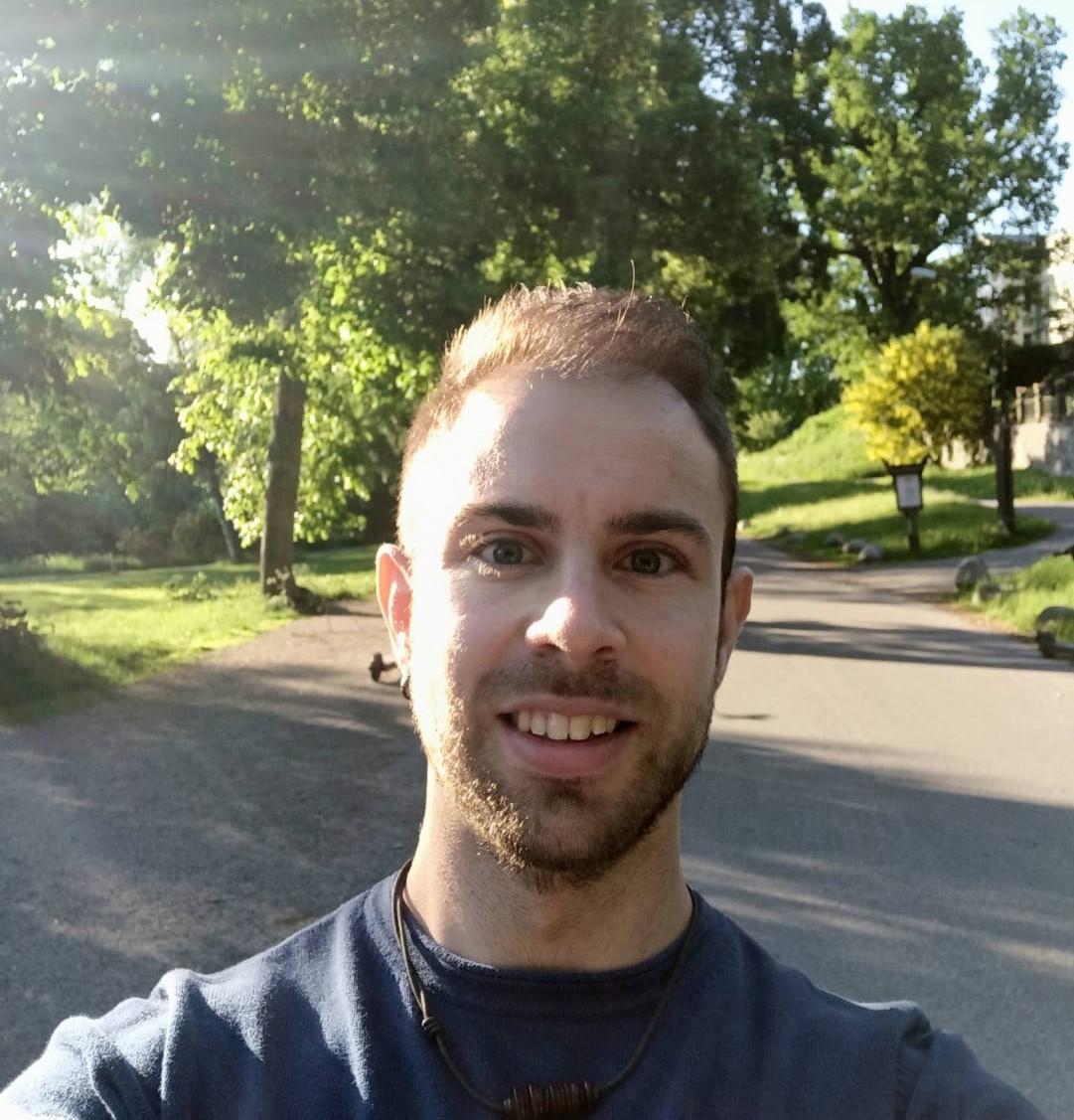}}

\begin{document}
\vspace*{4cm}
\title{Optimized FLUKA cross sections for cosmic-ray propagation studies}

\author{\textbf{P.~De~La~Torre~Luque$^{a}$}, M.~N.~Mazziotta$^{b}$}

\address{$^{a}$ The Oskar Klein Centre, Department of Physics, Stockholm University, AlbaNova\\
  SE-10691 Stockholm, Sweden
\\ \noindent $^{b}$Istituto Nazionale di Fisica Nucleare, Sezione di Bari, via Orabona 4, I-70126 Bari, Italy}

\maketitle\abstracts{
The current great precision on cosmic-ray (CR) spectral data allows us to precisely test our simple models on propagation of charged particles in the Galaxy. However, our studies are severely limited by the uncertainties related to cross sections for CR interactions. Therefore we have developed a new set of cross sections derived from the {\tt FLUKA} Monte Carlo code, which is optimized for the treatment of CR interactions. We show these cross sections and the main results on their application for CR propagation studies. Finally, we discuss the prediction of a low-energy break in the electrons spectra inferred from gamma-ray data.
}

\section{Introduction}

Galactic CRs, injected and accelerated in astrophysical sources, propagate throughout the Galaxy for millions of years, occasionally interacting with the gas in the interstellar medium (ISM) through spallation reactions that produce secondary particles, such as gamma rays, neutrinos or secondary nuclei that we call secondary CRs. The amount of secondary CRs formed provides direct information about the mean grammage traversed by the so-called primary CRs (i.e. the amount of gas per unit area that primary CRs cross during their journey) and, therefore, about the time that these CRs reside in the Galaxy. Measuring the spectra of secondary CRs such as Li, B, Be or F provides further information on how CRs interact with the magneto-hydrodynamical (MHD) turbulence in the ISM plasma and the energy dependence of these interactions.

The transport of CRs in the Galaxy is conventionally studied as a diffusive process characterised by a diffusion coefficient which is, basically, a power-law in energy ($D \propto E^{\delta} $) and whose spectral index $\delta$ is intimately related to the MHD interactions that CRs undergo. The ratios of the fluxes of secondary CRs to primary CRs, such as the B/C ratio, are used to determine the diffusion coefficient, given that they are directly related to each other, as can be seen from the approximate relation $N_{sec}/N_{prim}(E) \propto \sigma(E)/D(E)$, where $\sigma(E)$ is the spallation cross section of production of the secondary CR involved. Therefore, a precise evaluation of these ratios is crucial nowadays to unveil the process of transport of charged particles in the Galaxy and its main features, which also requires a good knowledge of cross sections of CR interactions in a broad energy range.
However, while current CR data is extremely precise since the past decade (at the level of a few percent), recent studies have proved the need of improving cross sections on spallation reactions in order to reduce the uncertainty related to the determination of the diffusion coefficient from the secondary-to-primary flux ratios~\cite{Luque:2021joz,Weinrich_combined,Korsmeier:2021brc}. In fact, the accuracy of spallation cross sections measurements for the energy range between the GeV and the TeV is very poor (20\%) and frequently there is not data at all for the energies of interest. On top of this, given that the nuclear models describing these kind of interactions are mainly adjusted to reproduce accelerator data, the cross sections computed from fundamental models are not totally consistent with the available data in the GeV-TeV range and, hence, we usually rely on parameterisations, fitted to the very scarce, limited and uncertain experimental data, instead. 

Nevertheless, these simulation codes, based on Monte Carlo simulations and event generators experienced a positive boost in the last years, usually driven by radiological and medical applications, which need to accurately describe the transport of ions in different materials and their interactions. Recently, the FLUKA Monte Carlo nuclear code~\footnote{http://www.fluka.org/fluka.php} has been optimized to be used in different kinds of astrophysical studies, with special attention to CR interactions~\cite{FLUKA,Mazziotta:2020uey,Mazziotta:2015uba,Fermi-LAT:2016tkg}. {\tt FLUKA} is a general purpose tool that can be used to transport particles in arbitrarily complex geometries and magnetic fields and study their nuclear interactions with hadrons and nuclei from the MeV/n up to $10$~PeV. %(mega-electronvolt per nucleon)

Here, we overview the main results presented in~\textit{De la Torre et al., (2022)}~\cite{Luque:2022aio}, where the full cross-section network for CR interactions up to iron nuclei (Z=26) was computed using {\tt FLUKA}. These cross sections are tested against data and implemented in a customised version~\footnote{A similar version of this code can be downloaded at \url{https://doi.org/10.5281/zenodo.4461732}} of the DRAGON2 code~\cite{DRAGON2-1,DRAGON2-2} in order to study the different ratios of the secondary CRs B, Be and Li and the propagation parameters derived from these predictions through a Markov chain Monte Carlo (MCMC) analysis~\cite{Luque_MCMC} of the most recent data from the AMS-02 collaboration~\cite{AMS02_BBeLi}.
Furthermore, we demonstrate that gamma-ray data indicate the need of including a break at low energies ($< 10$~GeV) in the injection spectrum of electrons.

\section{The derived FLUKA cross sections}
Inelastic and inclusive cross sections of all stable isotopes from protons to iron impinging on helium and hydrogen as targets (representing the composition of the ISM gas,) from $1$~{MeV/n} to $35$~{TeV/n}, are calculated with {\tt FLUKA} using 176 bins equally spaced in a logarithmic scale.

Inelastic cross sections are those related to the probability of destruction of a nucleus when it interacts with another particle and its effects in the spectrum of CRs become only significant at low energies, when the time-scale of inelastic collisions ($\tau_{inel}^{-1} \sim v n_{ISM} \sigma_{inel}(E)$, where $v$ is the speed of the particle, $n_{ISM}$ is the target number density and $\sigma_{inel}$ is the inelastic cross section of the interaction) is smaller than the diffusion time-scale ($\tau \sim H^2/2D(E)$, where $H$ is Galactic halo size and $D(E)$ is the diffusion coefficient). Experimental data on inelastic cross sections with proton as target are measured with a precision $\leq 15\%$ in the GeV range. The inelastic cross sections computed with {\tt FLUKA} show a good agreement with data and with other dedicated parameterisations, consistent in all the energy range studied within $\sim25\%$ discrepancies. For interactions of nuclei heavier than Ne (Z=10) with protons, for which experimental data in the GeV range is mainly absent, the cross sections predicted from {\tt FLUKA} differ from those predicted from dedicated parameterisations by a $\sim20\%$ factor, roughly constant in energy~\cite{Luque:2022aio}.

In turn, inclusive cross sections are that fraction of the inelastic cross sections which go into the creation of a secondary nucleus and regulate the rate of production of secondary CRs ($\tau_{p\rightarrow s}^{-1} \sim v_p n_{ISM}\sigma_{{p\rightarrow s}}(E)$, where $v_p$ is the speed of the projectile nucleus, p, and $\sigma_{{p\rightarrow s}}$ is the cross section of production of the secondary particle, s, from the interaction of the projectile particle with the target nucleus). These are poorly known because of the experimental difficulties to perform the measurements and, therefore, the uncertainties associated to inclusive cross-section data are of $\sim 20\%-30\%$ in the GeV range. On top of this, inclusive cross sections should also include the decay of ghost nuclei (i.e. short-lived nuclei generated in spallation reactions, which have a negligible lifetime compared to typical CR propagation times, and decay into the secondary nucleus that we are considering) in the current CR propagation codes. The inclusive cross sections including this effect are referred to as cumulative inclusive cross sections.
\begin{figure}[t!]
\includegraphics[width=0.35\linewidth]{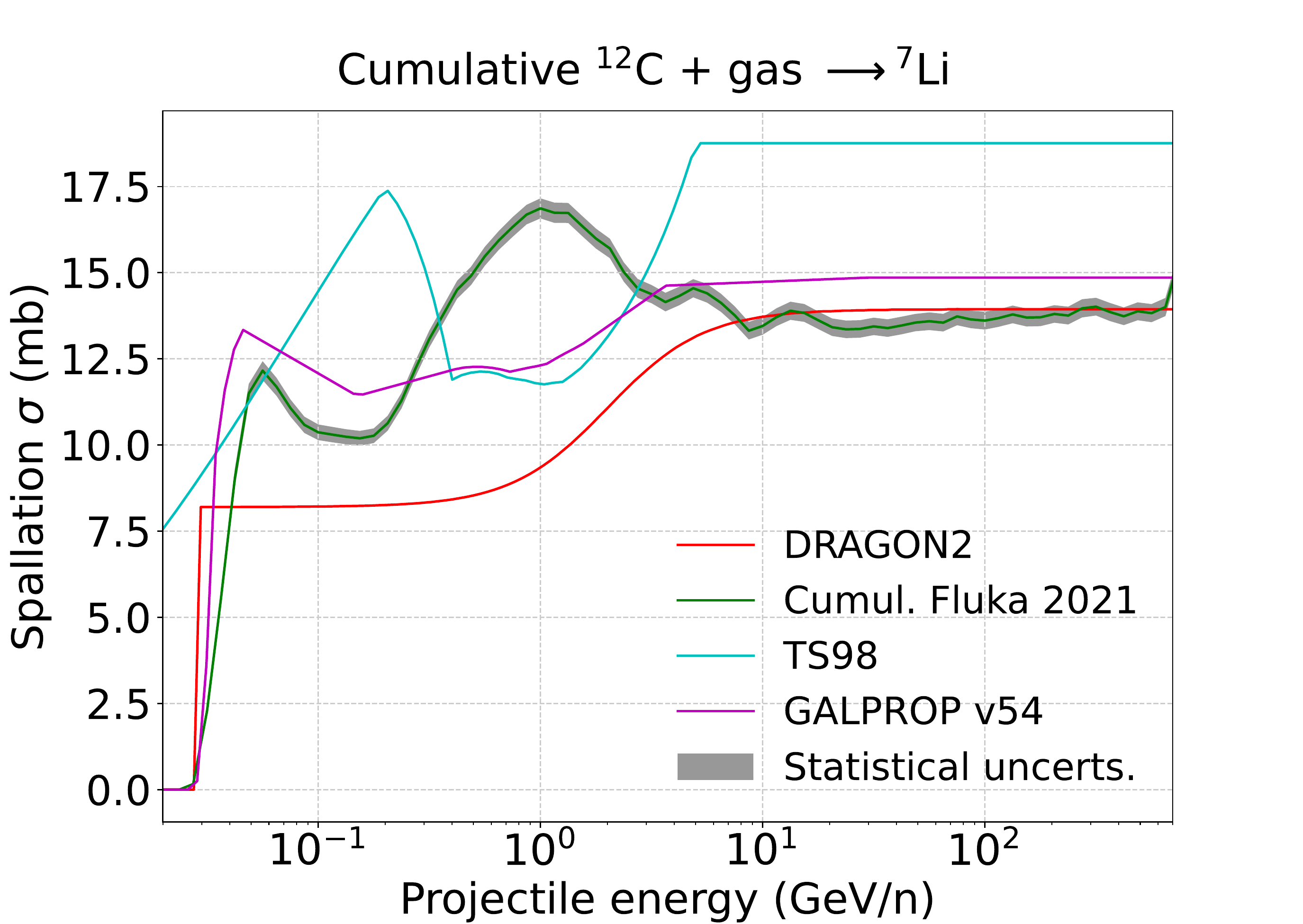} \hspace{-0.68cm}
\includegraphics[width=0.35\linewidth]{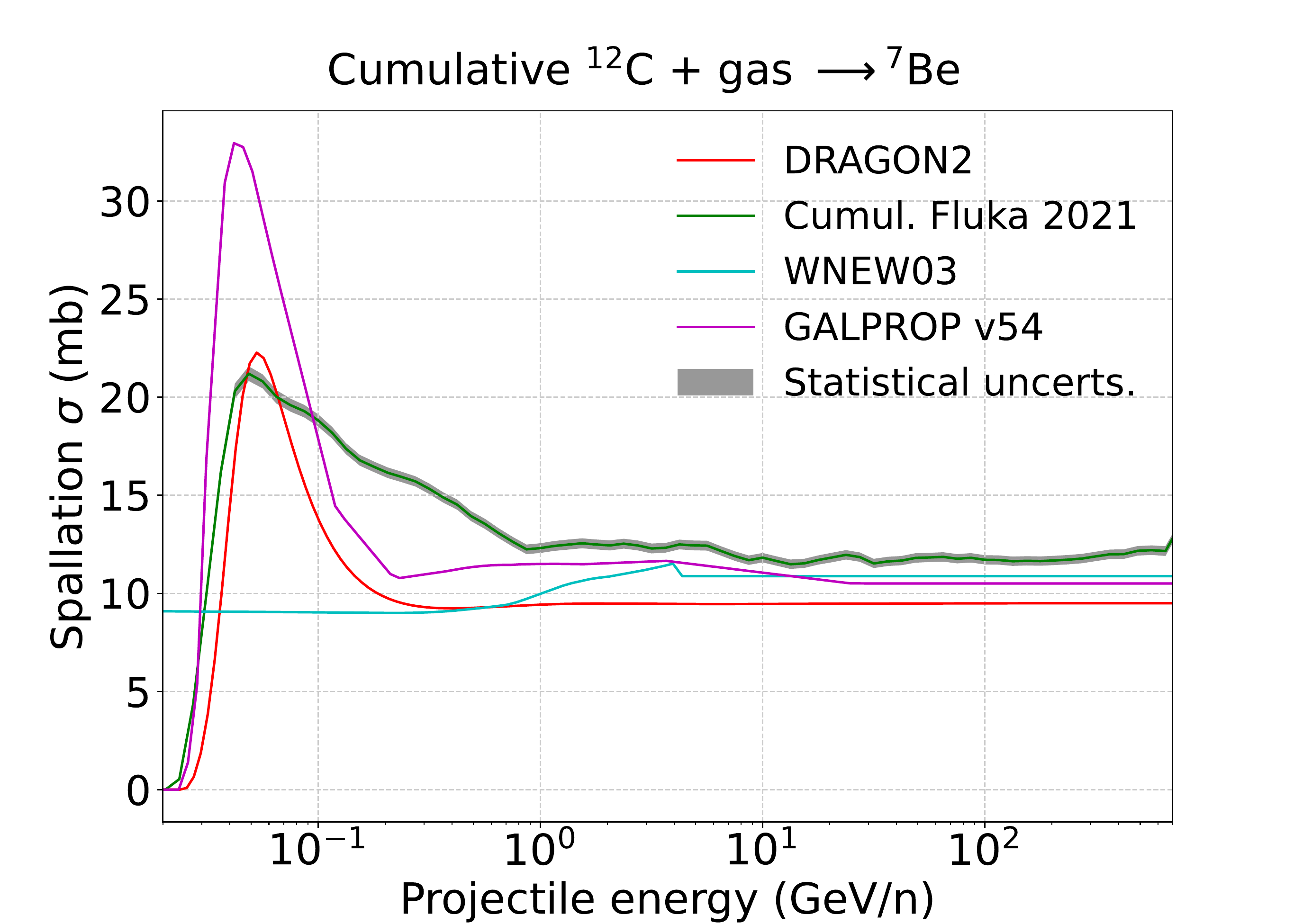} \hspace{-0.68cm}
\includegraphics[width=0.35\linewidth]{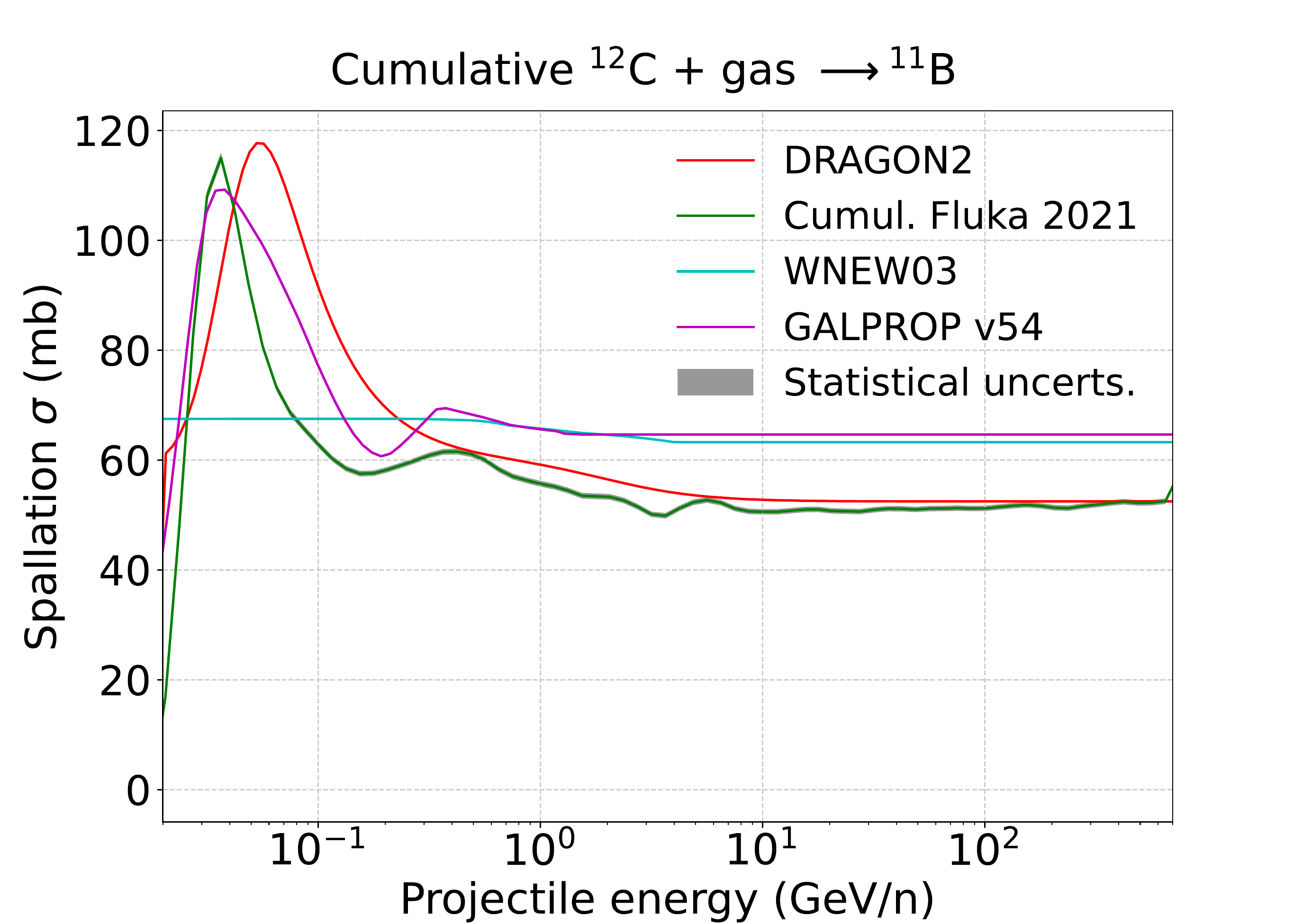}

\includegraphics[width=0.35\linewidth]{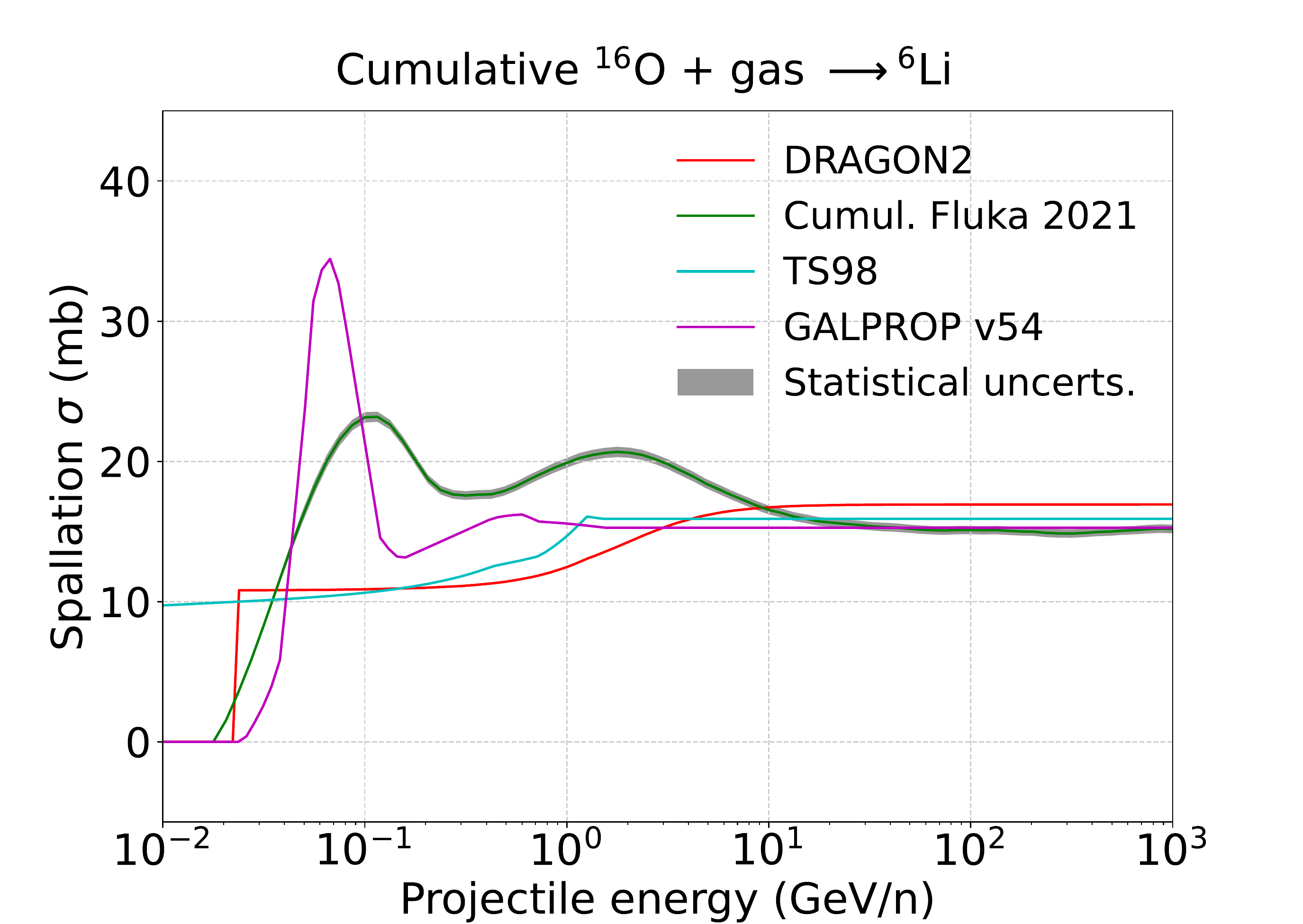} \hspace{-0.68cm}
\includegraphics[width=0.35\linewidth]{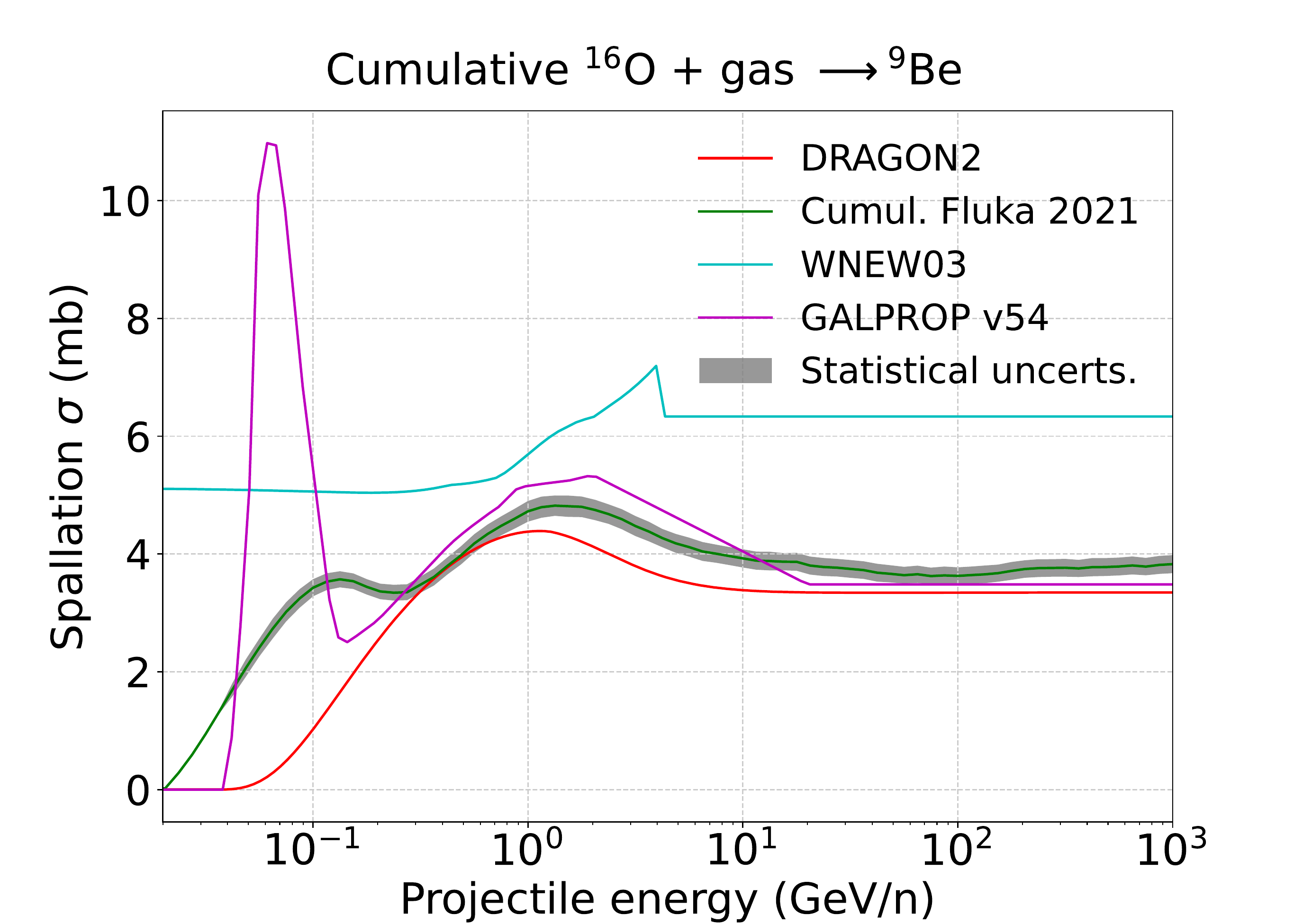} \hspace{-0.68cm}
\includegraphics[width=0.35\linewidth]{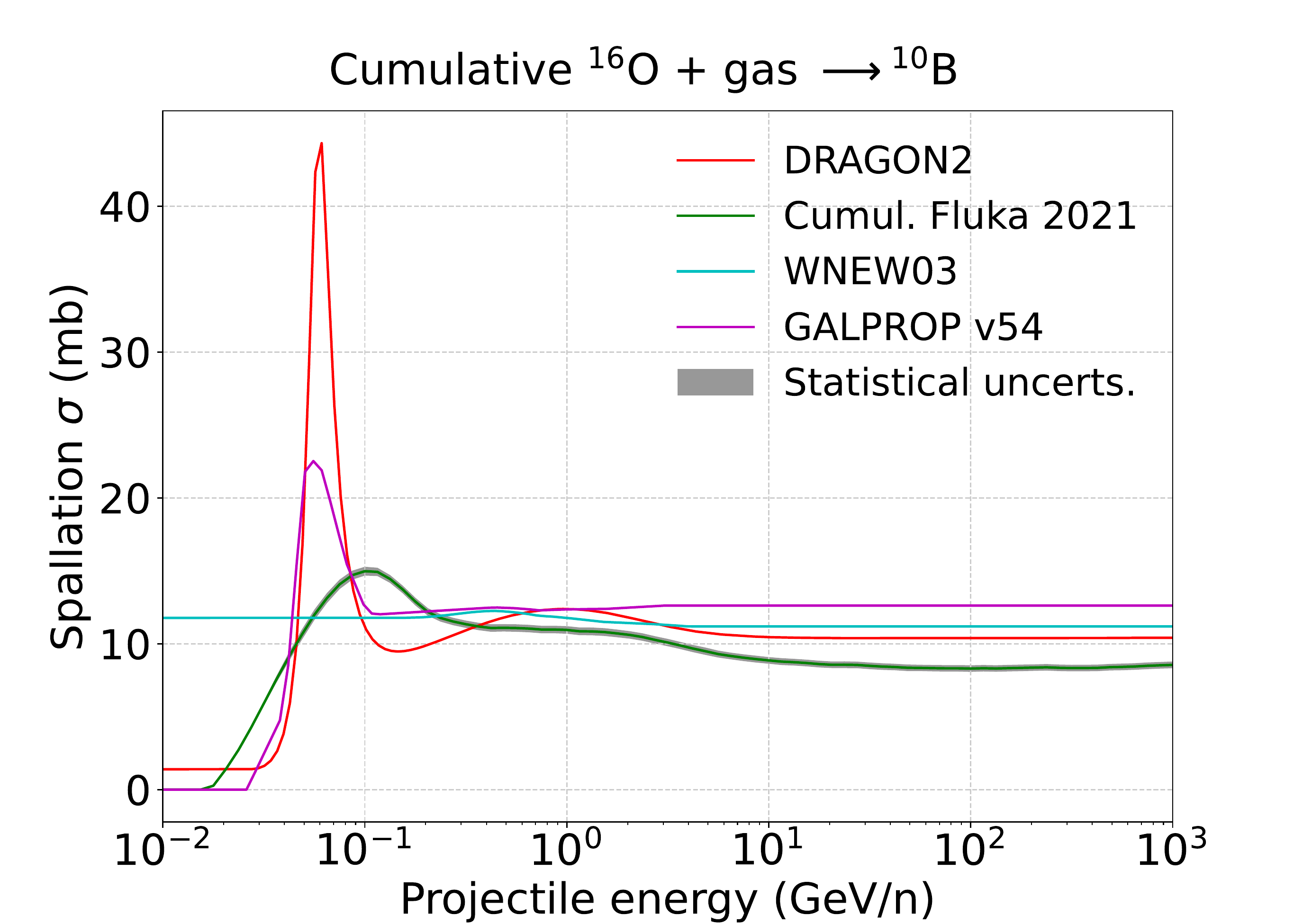}
\caption{Spallation cross sections of CRs interaction with ISM gas computed with {\tt FLUKA} compared to the most widely used parameterisations, for the production of isotopes of B, Be and Li from $^{12}$C and $^{16}$O as projectiles.}
\label{fig:Cumul_XS}
\end{figure}
Figure~\ref{fig:Cumul_XS} shows the total (i.e. cross sections of interactions with a gas with ISM composition) cumulative cross sections of production of the some isotopes of B, Be and Li from C and O (the main producers of these secondary CRs). Here, we compare the computed {\tt FLUKA} cross sections with well-known dedicated cross sections parameterisations, commonly used in CR codes: the GALPROP~\cite{GALPROPXS} and DRAGON2~\cite{Luque:2021joz} parameterisations and the cross sections calculated with the {\tt WNEW03}~\cite{webber2003updated} and {\tt YIELDX}~\cite{silberberg1998updated} (TS98 in the legend) codes. 
As we see, the FLUKA cross sections are very consistent with the predictions from the other parameterisations, which is remarkable since our predictions completely rely on theory-based interaction models and not in fits to cross sections data. We highlight that both the normalization and energy dependence of these cross sections is compatible with the most updated parameterisations in the whole energy range. In addition, the position of the predicted resonances are also in good agreement with those parameterised from data. 
We observe that such good consistency is found in most channels of production of isotopes of B, Be and Li from interactions of C, N, O, Ne, Mg and Si with gas, which constitute the main primary CRs producing these secondary CRs. However, we observe that the channels of production of F (produced mainly from Ne, Mg and Si) predict much lower cross sections (by a factor of $\sim2$), which seem to be not compatible with the few existent data. Finally, we remark the importance of having reliable cross sections predictions for those channels in which experimental data is totally absent (meaning that the parameterisations are merely extrapolations in these channels), as happens for the channels production of heavy secondary cosmic rays, like $^{26}$Al, for example.

\section{CR transport predictions using the FLUKA cross sections}
To study the spectra of B, Be and Li derived with the FLUKA cross sections and the propagation parameters inferred from their ratios, we implement these cross sections in the DRAGON2 code.
The same injection and propagation set-up as presented our previous works~\cite{Luque:2021joz,Luque_MCMC,Luque:2022aio} is used, which includes reacceleration and neglects convection. Solar modulation is treated with the Force-field approximation~\cite{FF}. In this work, we focus on the results obtained with the diffusion coefficient parameterised as in Equation~\ref{eq:breakhyp}. In this equation, the free parameters entering the fit are the normalization, $D_0$, the exponent of $\beta$ (speed of particle in units of the speed of light), $\eta$, and the spectral index, $\delta$, while $R_0$, the rigidity at which the diffusion coefficient is normalized, is set to $4$~GV. Then, we fix $\Delta\delta = 0.14$, $R_b = 312$~GV and $s = 0.04$~\cite{Luque_MCMC}. The use of the diffusion coefficient of Eq.~\ref{eq:breakhyp} requires using as injection a broken power-law with a break set at $8$~GeV.
\begin{equation}
 D = D_0 \beta^{\eta}\frac{\left(R/R_0 \right)^{\delta}}{\left[1 + \left(R/R_b\right)^{\Delta \delta / s}\right]^s} \,\,\, ,
\label{eq:breakhyp}
\end{equation}

The starting point of this analysis consist of finding the propagation parameters that allow us to reproduce the boron-over-carbon (B/C) spectrum reported by AMS-02. The injection spectra of nuclei up to iron are adjusted, recursively, along with the propagation parameters, to reproduce AMS-02 data~\cite{AMS02_BBeLi}. This analysis yielded values of the propagation parameters entering in the fit which are reported in Table~\ref{tab:BC_params}. This table also reports those parameters found for the same analysis for the DRAGON2 and GALPROP parameterisations, for comparisons. As we can see, the parameters obtained using the FLUKA cross sections are consistent within $1\,\sigma$ with those obtained using dedicated parameterisations, except for the normalization of the diffusion coefficient. This is something remarkable and achieved for the first time with a set of cross sections derived from fundamental models of nuclear interactions.
\begin{table}[!t]
\centering
\begin{tabular}{|l|c|c|c|}
  \multicolumn{4}{c}{\hspace{0.3cm}\large} \\ \hline \textbf{B/C best-fit parameters} & \textbf{FLUKA}  & \hspace{0.2 cm}\textbf{GALPROP} & \hspace{0.2 cm}\textbf{DRAGON2}\\ 
  \hline
{D$_0$/H} ($10^{28}$ cm$^{2}$ s$^{-1}$ kpc$^{-1}$) & 0.82 $\pm$ 0.03 & 0.94 $\pm$ 0.04 & 0.97 $\pm$ 0.04\\
{$v_A$} (km/s) & 23.3 $\pm$ 2.3 & 24.4 $\pm$ 3. & 22. $\pm$ 3.6\\
{$\eta$} & -0.67 $\pm$ 0.13 & -0.66 $\pm$ 0.12 & -0.78 $\pm$ 0.13\\      
{$\delta$}  & 0.45 $\pm$ 0.01 & 0.45 $\pm$ 0.01 & 0.44 $\pm$ 0.01\\   
\hline
\end{tabular}
\label{tab:BC_params}
\caption{Propagation parameteres found in the MCMC analysis of the B/C flux ratio. The halo size value, H, obtained from the $^{10}$Be flux ratios is $7.54$~kpc. The errors shown here correspond only to statistical uncertainties.}
\end{table}

To evaluate the size of the magnetised halo we analyse the ratios of $^{10}$Be to $^{9}$Be and to the total flux of Be. Full details can be found in our reference work~\cite{Luque:2022aio}. This fit yields a value of the halo size of $\sim7.5^{+1.13}_{-0.95}$~kpc, similar to estimations with other cross sections ($H\sim6.76\pm1$~kpc for the DRAGON2 cross sections and $H\sim6.93\pm0.98$~kpc for the GALPROP ones).

One of the most important tools for the study of the cross sections used to evaluate the spectra of secondary CRs is the flux ratios among them (Be/B, Li/B and Li/Be), which are shown in Fig.~\ref{fig:Ratios_Fluka}. At high energy, these ratios mainly depend on the ratio of the cross sections of the CRs involved and the injection spectra of primary CRs, which are adjusted to reproduce AMS-02 data. In this figure, we show a band of statistical uncertainty associated to the uncertainty related to the determination of the FLUKA cross sections. As can be seen in the figure, while the Li/Be ratio reproduces within $1\,\sigma$ uncertainty AMS-02 data above $\sim2$~GeV, the ratios involving B are overproduced by a roughly constant $20-25\%$ above $3$~GeV. We remark that this discrepancy is yet within the typical uncertainty on cross sections measurements above the GeV. However, this discrepancy means that, as happens with the parameterisations commonly used in CR studies, they are not able to reproduce simultaneously B, Be and Li. A simultaneous solution to this discrepancy and the discrepancy found in the value of $D_0$ with respect to the parametrisations would be to renormalise (to lower) the cross sections of B production by around a $20\%$. The result of applying this scaling to the B production cross sections is shown in the figure as a dashed line and leads to a good simultaneous reproduction of the AMS-02 spectra of B, Be and Li above $\sim3$~GeV.
\begin{figure}
\begin{minipage}{0.333\linewidth}
\includegraphics[width=1.1\linewidth, height=4.cm]{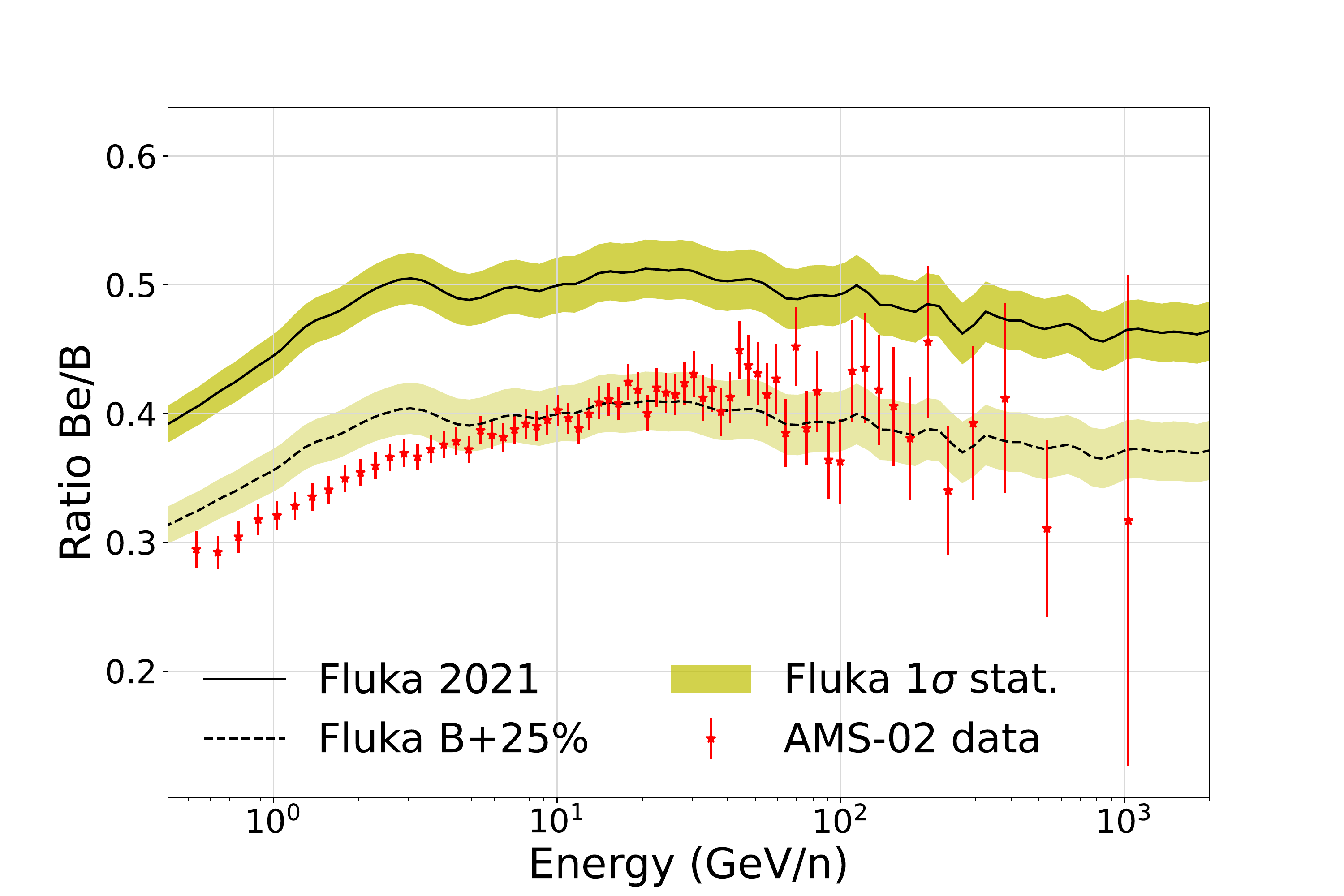}
\end{minipage}
\begin{minipage}{0.323\linewidth}
\includegraphics[width=1.1\linewidth, height=4.cm]{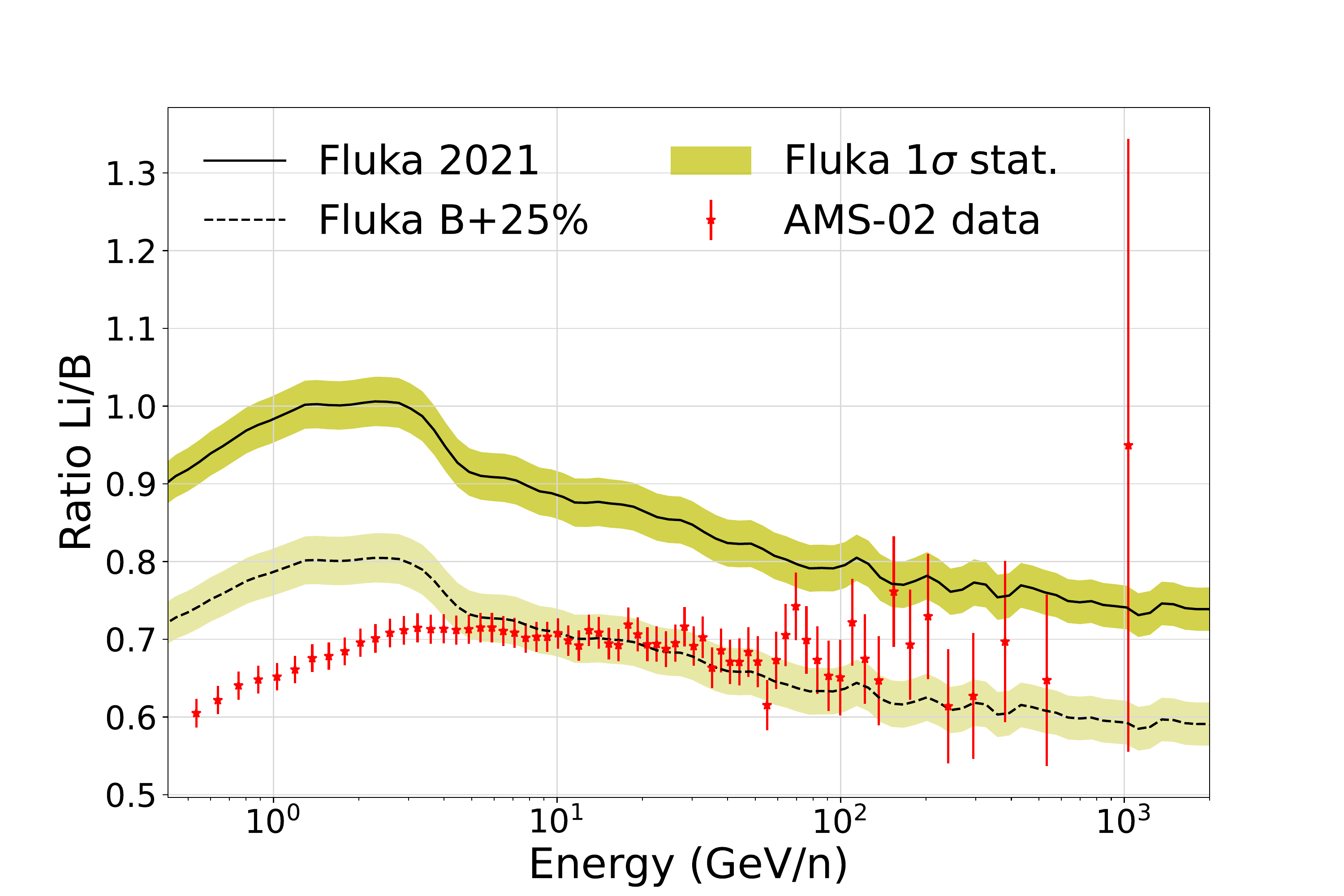}
\end{minipage}
\begin{minipage}{0.333\linewidth}
\includegraphics[width=1.1\linewidth, height=4.cm]{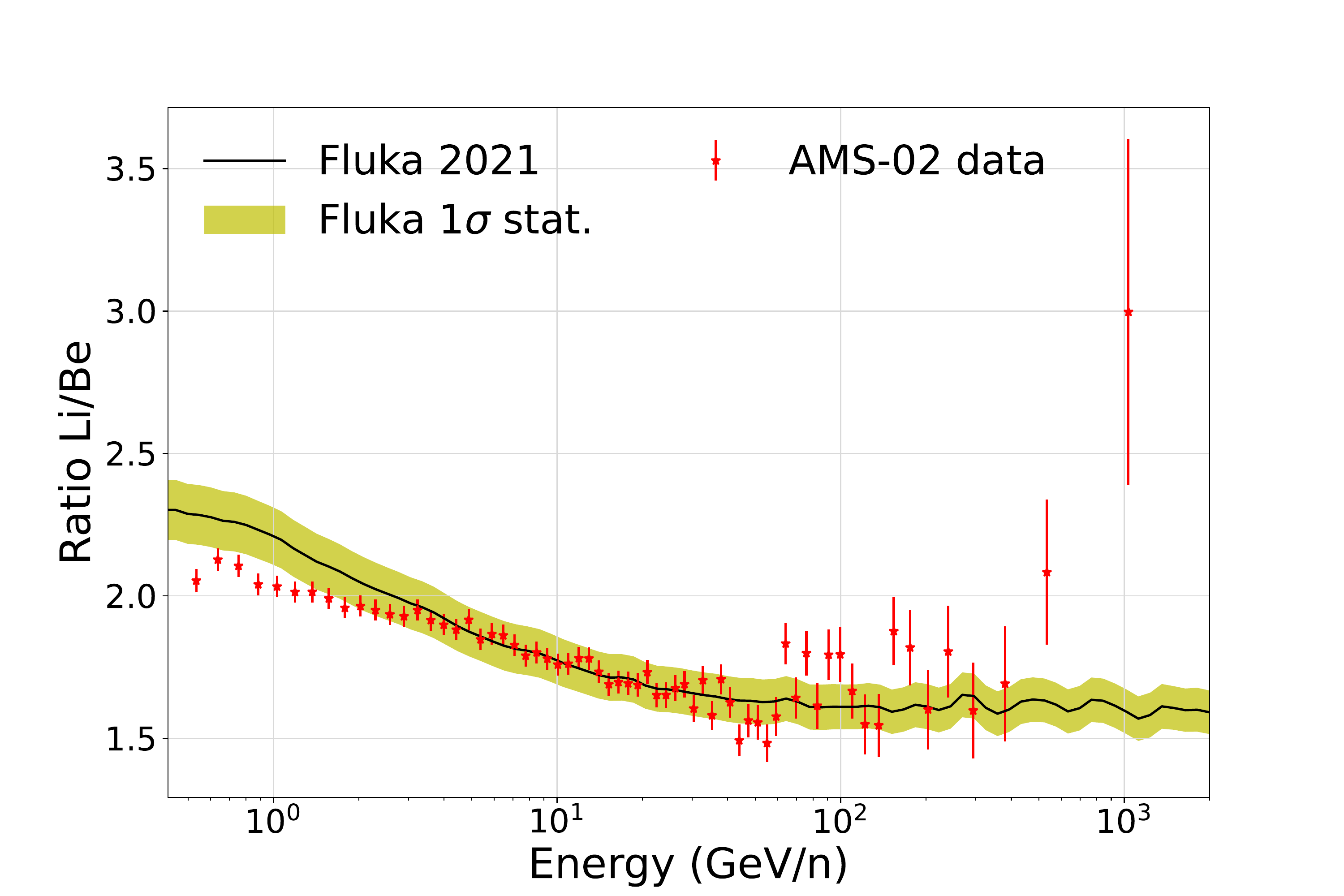}
\end{minipage}
\label{fig:secsec_Fluka}
\caption{Be/B, Li/B and Li/Be ratios obtained with the FLUKA computations, together with the band of statistical uncertainties due to the spallation cross sections calculation (since this is a Monte Carlo computation).}
\end{figure}

\subsection{Combined analysis of B, Be and Li}
\begin{figure}[t!]
\includegraphics[width=0.5\linewidth]{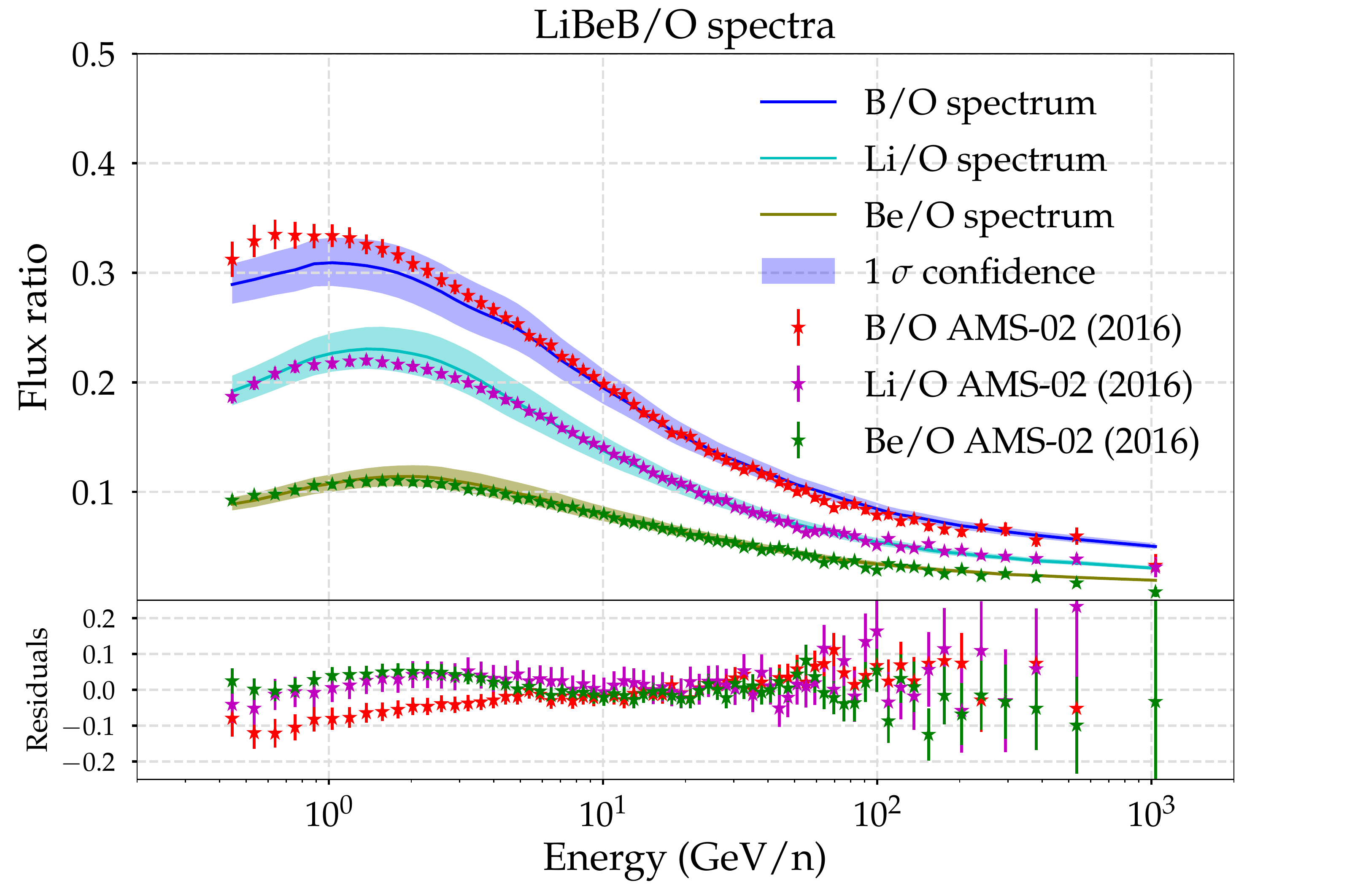}
%\hspace{0.13cm}
\includegraphics[width=0.5\linewidth]{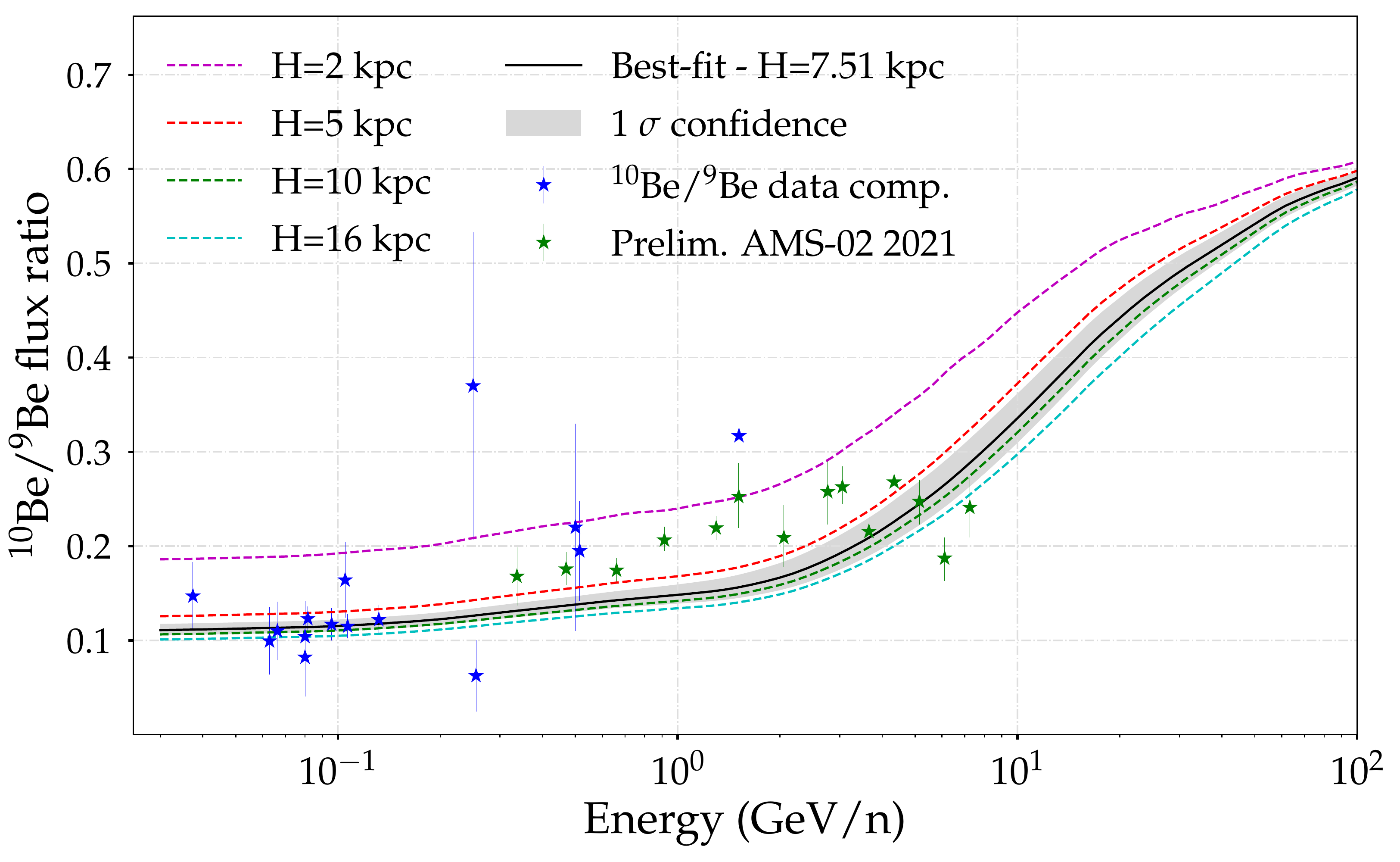}

\includegraphics[width=0.5\linewidth]{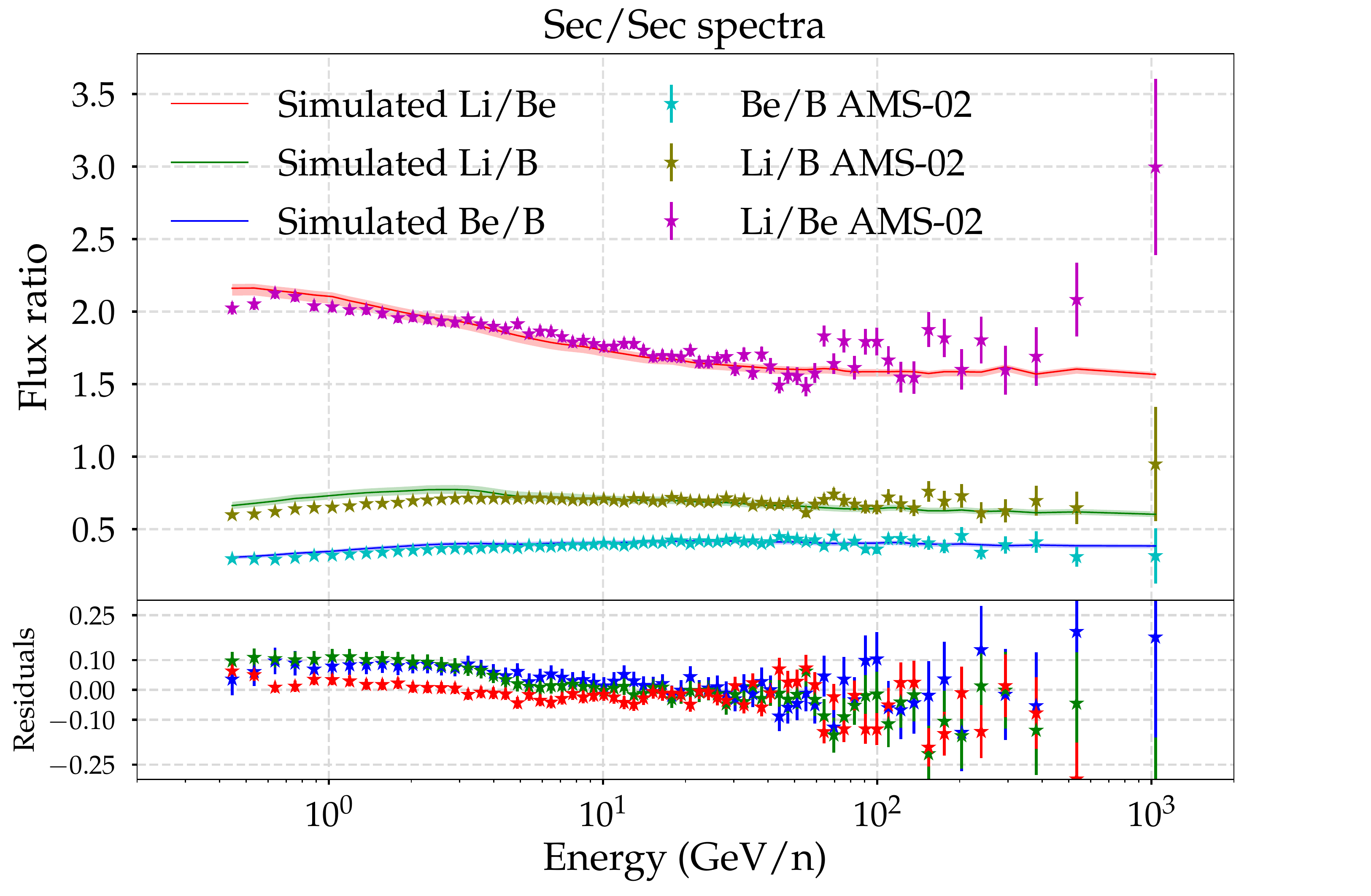}
%\hspace{0.1cm}
\includegraphics[width=0.5\linewidth]{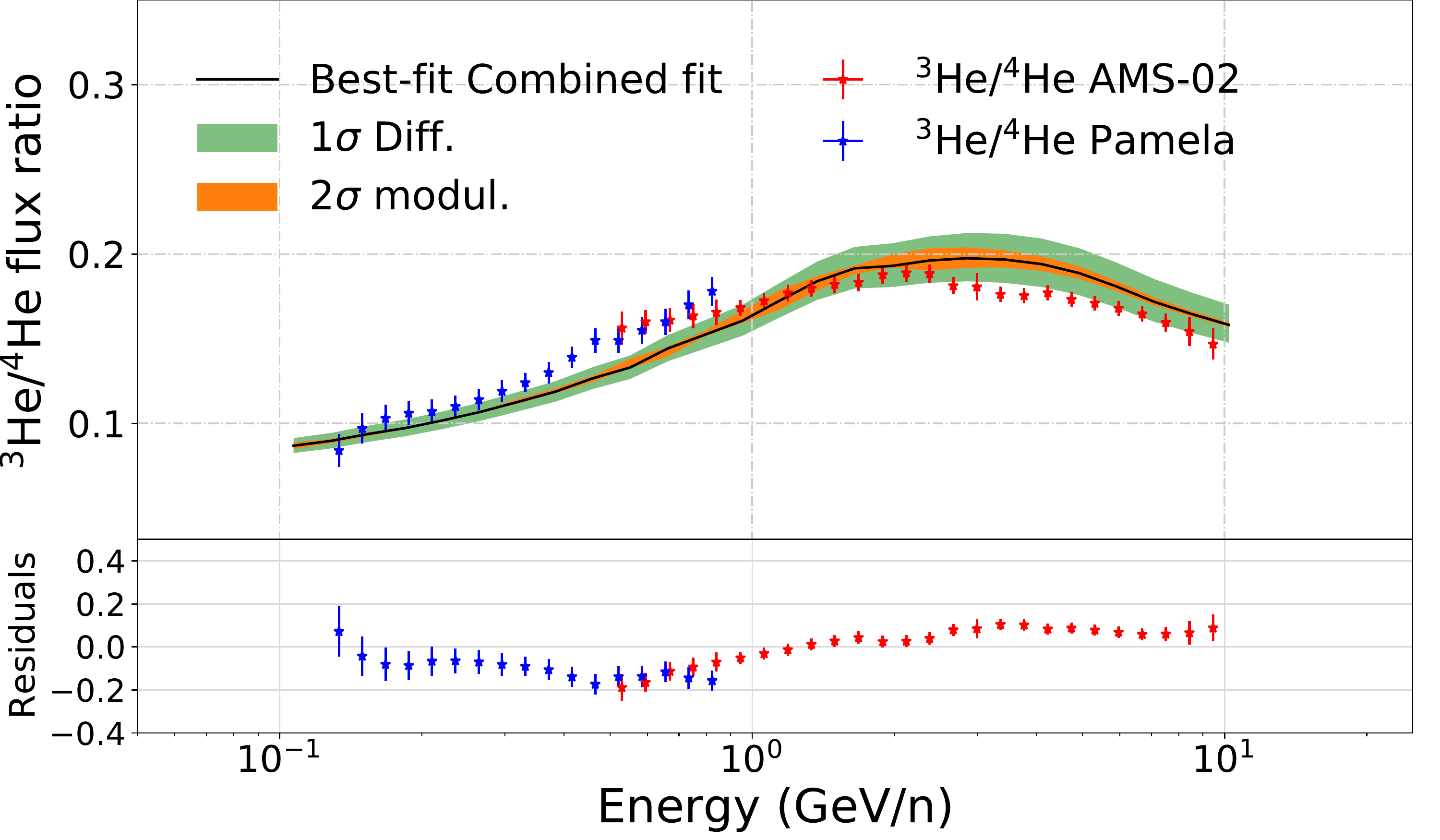}
\caption{\textbf{Top-left panel:} Secondary-over-secondary flux ratios of B, Be and Li computed using the propagation parameters and cross sections scaling factors found in the combined analysis, compared to the AMS-02 experimental data. \textbf{Lower-left panel:} B/O, Be/O and Li/O flux ratios predicted with the parameters determined by the MCMC combined analysis for the DRAGON2 cross sections. 
\textbf{Top-right panel:} $^{10}$Be/$^{9}$Be flux ratio compared to available experimental data.
\textbf{Lower-right panel:} Predicted $^3$He/$^4$He flux ratio compared to Pamela and AMS-02 data. Residuals, calculated as model-data/model are also shown. This ratio is calculated with the diffusion coefficient obtained from the combined fit.}
\label{fig:Ratios_Fluka}
\end{figure}

The final test in the study of the spectra of secondary CRs evaluated with the FLUKA cross sections is a combined analysis where we combine the ratios of B, Be and Li to C and O (B/C, B/O, Be/C, Be/O, Li/C and Li/O) and its respective secondary-to-secondary flux ratios (Be/B, Li/B, Li/Be), MCMC algorithm presented in a precedent work~\cite{Luque_MCMC}. This analysis includes scaling factors, as nuisance parameters, that renormalise the cross sections of production of B, Be and Li and, associated to this scaling factor, there is a penalty factor that penalises large variations of the original cross sections. The result of this analysis is a set of propagation parameters and scaling factors that are able to simultaneously reproduce all the ratios of B, Be and Li within the $1\,\sigma$ statistical uncertainties in the determination of the propagation parameters. This is shown in Figure~\ref{fig:Ratios_Fluka}, where we report the ratios of Be/B, Li/B and Li/Be, B/O, Be/O, Li/O and the $^{10}Be$/$^{9}$Be ratio, included in the fit. We highlight that the diffusion coefficient obtained from these combined analyses allows us to reproduce, at the same time, the $^3$He/$^4$He flux ratio, shown in the lower-right panel, which is evidence that the diffusion coefficient predicted by the light secondary CRs is compatible for the different nuclei within uncertainties. 

These analyses predict a spectral index of the diffusion coefficient $\sim 0.36$, which is significantly lower than the typical values found from B ratios, although compatible with the value found for the combined analysis with the GALPROP cross sections~\cite{Luque_MCMC}. Furthermore, this value is consistent with the basic predictions from wave-particle interactions, for which the standard spectrum of plasma waves leads to a spectral index of the diffusion coefficient which is $0.33 \lesssim \delta \lesssim 0.5$. Finally, another crucial point of this analysis is the cross-section scaling factors obtained. These are $1.18$, $0.94$ and $0.93$, for B, Be and Li, respectively, with $1\sigma$ statistical uncertainties of $\sim \pm 0.01$. However, we highlight that there are other systematic uncertainties related to the determination of these scale factors, like those related to the gas distribution used, and could make the total systematic uncertainties in the determination of these scale factors larger than $5\%$~\cite{Luque_MCMC}. 
%Finally....The best-fit propagation parameters and their $2\,\sigma$ confidence contours are shown in the form of a triangle plot in Figure~\ref{fig:Box_plots}.

\begin{comment}
\begin{figure}
\begin{minipage}{0.45\linewidth}
\includegraphics[width=1.\linewidth]{Joint_corners-FlukaGalpropDRAGON2_BBeLi_Diff.png}
\end{minipage}
\hspace{0.13cm}
\begin{minipage}{0.45\linewidth}
\includegraphics[width=1.\linewidth]{Joint_cornersScale-FlukaGalpropDRAGON2_BBeLi_Diff.png}
\end{minipage}
\label{fig:Box_plots}
\end{figure}
\end{comment}

\section{Evidence of a low-energy break in electron injection spectrum from gamma-ray data}
\begin{figure}[t]
\begin{center}
\includegraphics[width=0.6\linewidth]{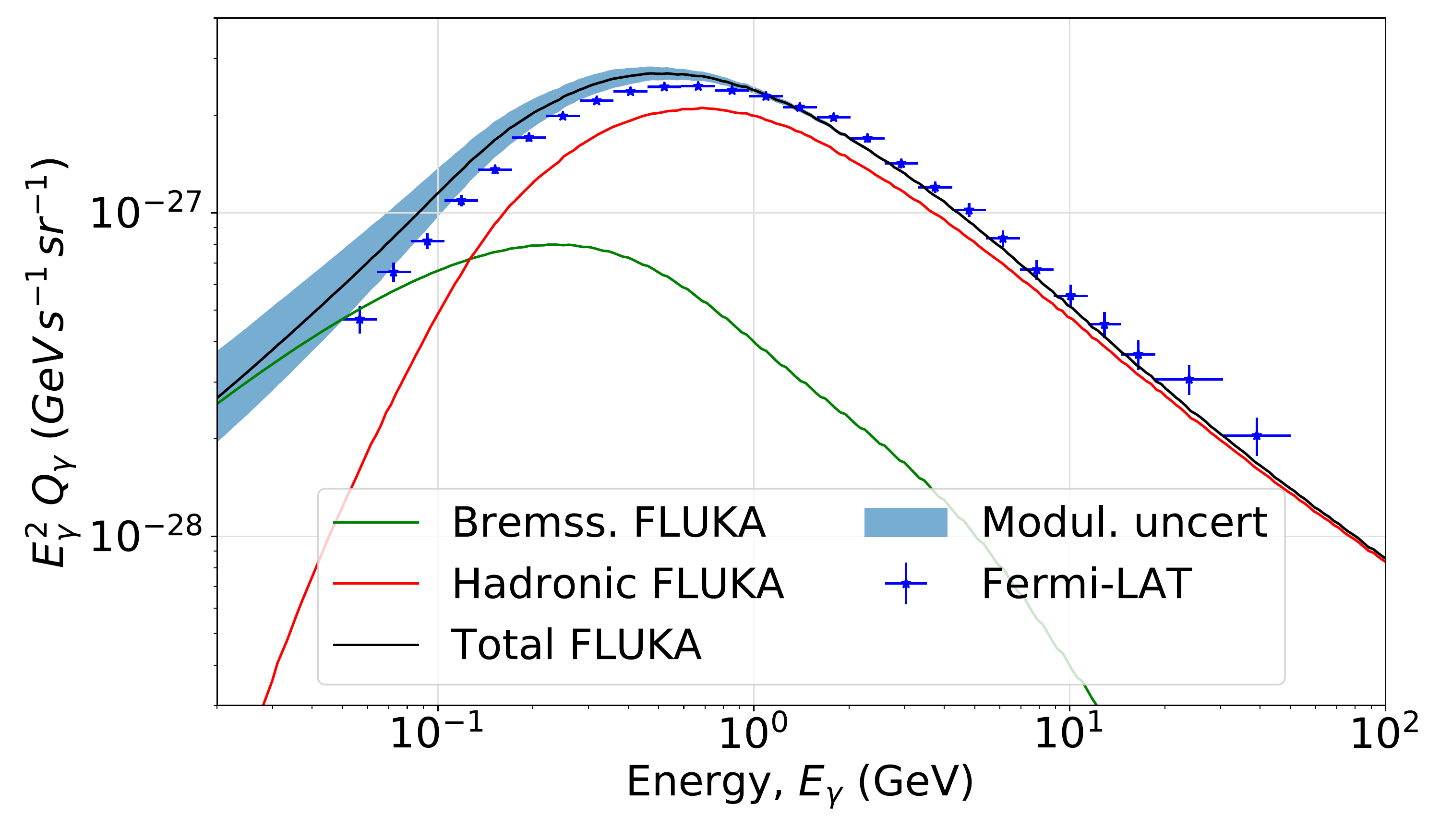}
\label{fig:Emiss}
\caption{Local HI gamma-ray emissivity spectrum for the propagation parameters derived from the combined analysis. An uncertainty band related to solar modulation uncertainty is also shown.}
\end{center}
\end{figure}

Finally, we employed the Local Interstellar spectrum (LIS) predicted from the combined analysis described above to study the diffuse gamma-ray emission, which is computed using the {\tt Gammasky} code
%~\cite{Cirelli:2014lwa,Gaggero:2015xza}. 
In particular, we tested the local HI gamma-ray emissivity spectrum using Fermi-LAT gamma-ray data~\cite{Casandjian:2015hja}. 
The dominant contribution to this observable at high energies is the $\gamma$-ray emission produced from the decay of unstable particles formed via nuclear reactions (hadronic emission), such as pions. As it is observed in Figure~\ref{fig:Emiss}, the hadronic emission allows us to reproduce the Fermi-LAT emissivity, within $1\,\sigma$ statistical uncertainties, assuming an ISM composition with relative abundance of H : He : C : N : O : Ne : Mg : Si = $1:0.096:4.65\times 10^{-4}:8.3\times 10^{-5}:8.3\times 10^{-4}:1.3\times 10^{-4}:3.9\times 10^{-5}:3.69\times 10^{-5}$.
%Our computation is performed using the gas maps developed by the GALPROP team~\cite{Moskalenko:2001ya,Ackermann:2012pya}. The distribution of sources is taken from Ref.~\cite{Ferriere2001}

At low energy ($E_{\gamma}<1$~GeV) bremsstrahlung emission from leptons becomes dominant. In order to evaluate the bremsstrahlung emission, we reproduce the AMS-02 e$^{+}+$e$^{-}$ data. We tested two different models that allow us to reproduce electrons: (i) a broken power-law with a break at $65$~GeV and a modified Force-field approximation, able to account for the polarity and charge sign of the different particles~\cite{Modified_FF}. (ii) a doubly-broken power-law with a break at $8$~GeV (same break position as it is commonly employed for nuclei) and a high energy break at $65$~GeV, using the conventional Force-field approximation~\cite{FF}. In this study, we found that the first option overproduces very significantly the local $gamma$-ray emissivity at low energies, while the second approach allows us to reproduce the emissivity within the uncertainties associated to the Fisk potential, assuming to be of $\phi \pm 0.1$, motivated by neutron monitors~\cite{Ghelfi}. 

\begin{figure}[b!]
\includegraphics[width=0.5\linewidth]{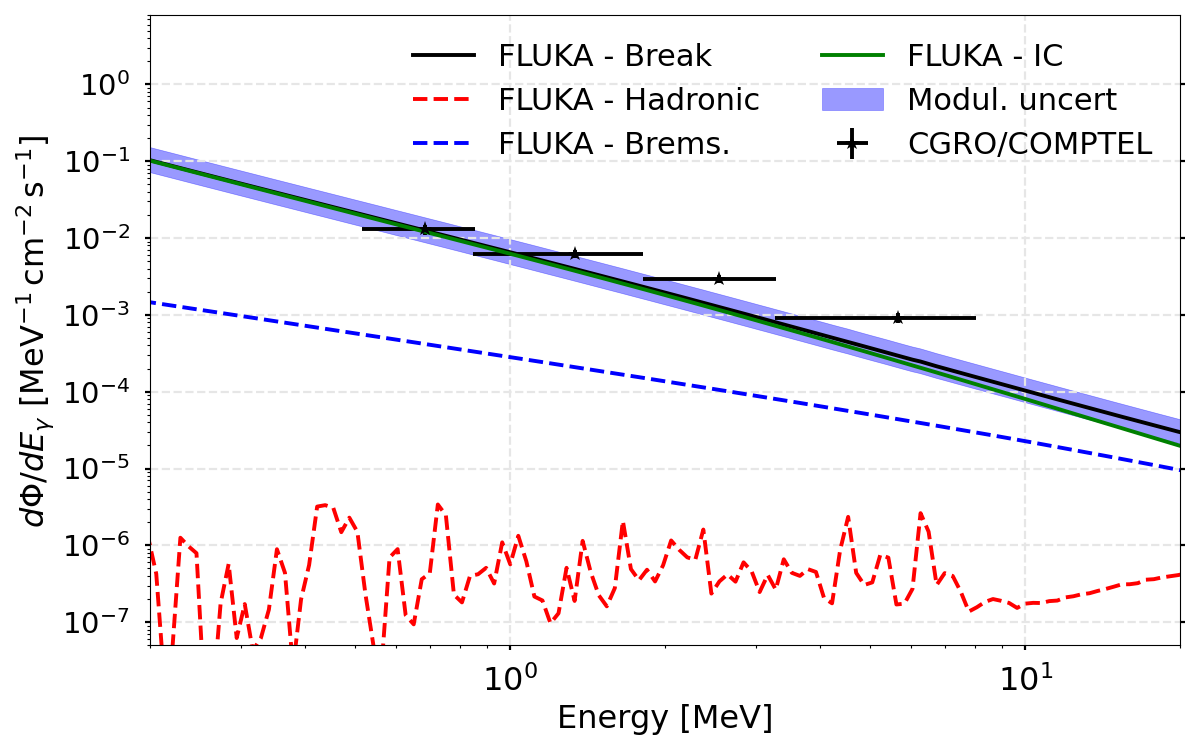}
\includegraphics[width=0.5\linewidth]{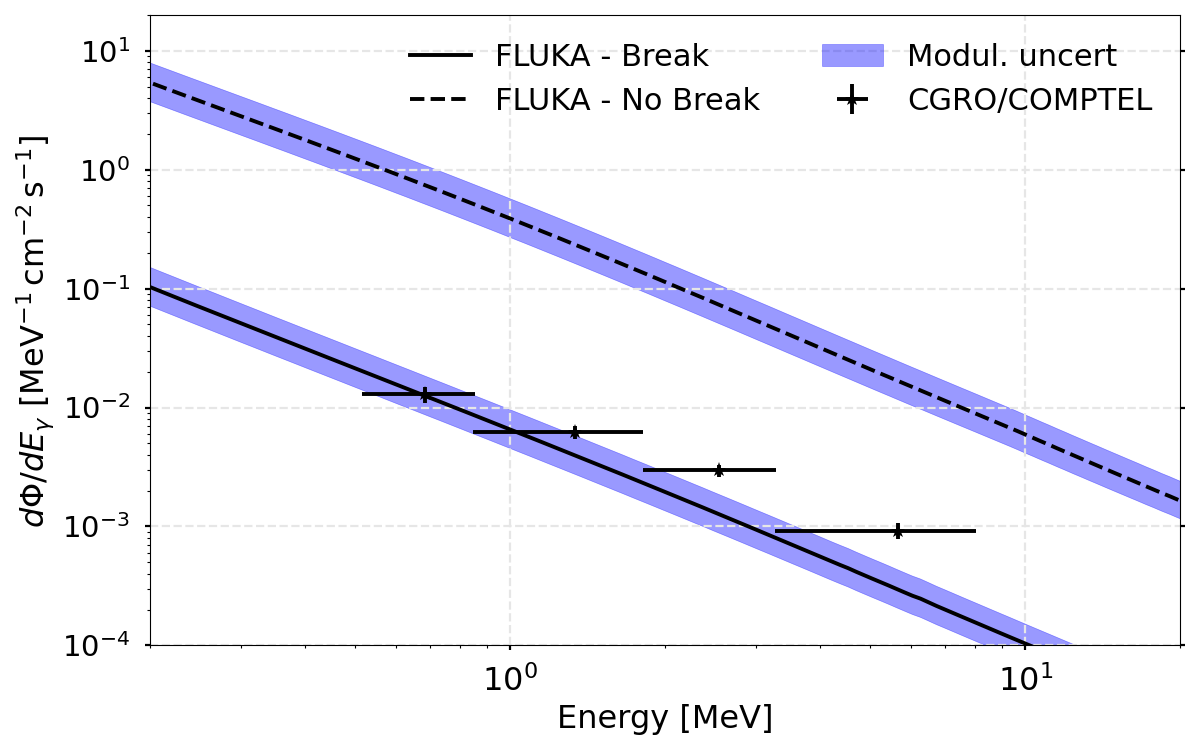}
\label{fig:MeVIC}
\caption{Predicted diffuse gamma-ray emission in the Galactic region $|b| < 47.5^{\circ}, |l| < 47.5^{\circ}$. The left panel shows the different contributions to the diffuse gamma-ray emission at those energies, while the right panel shows the predicted diffuse gamma-ray emission for the hypothesis of a low-energy break in the electron injection spectrum compared to the hypothesis without break.}
\end{figure}

This break was observed from synchrotron radio emission~\cite{Bernardo_2013} but never proved to be necessary to reproduce gamma-ray data, as we determine here. To further test the hypothesis of a break at the few GeV range in the electron spectrum we studied the diffuse gamma-ray emission at MeV energies around the center of the Galaxy ($|b| < 47.5^{\circ}, |l| < 47.5^{\circ}$), where Inverse-Compton (IC) emission produced from leptons interacting with the interstellar radiation fields is totally dominant, as it can be seen from Figure~\ref{fig:MeVIC} (left), where we show the predicted gamma-ray emission around the MeV for the model including a break at $8$~GeV in the electron spectrum. Hadronic emission in this plot corresponds to the gamma-ray emission from exited states of nuclei, computed using the FLUKA cross sections. As it can be seen, our predictions reproduce well the experimental data from COMPTEL telescope~\cite{CGRO}. Figure~\ref{fig:MeVIC} right shows a comparison between the predictions with and without a break in the electron spectrum, clearly supporting the break hypothesis, as the local emissivity does too.

\section{Conclusions}
\label{sec:conclusion}
In this work, we presented the FLUKA cross sections for Galactic CR propagation studies as an alternative and an improvement with respect to dedicated cross sections parameterisations.
We have tested the spectra of B, Be and Li and evaluated the main propagation parameters inferred from AMS-02 data, demonstrating that every CR observable can be reproduced with accuracy and at the level of precision of current CR parameterisations.

We showed that the FLUKA cross sections allow us to reproduce simultaneously the light secondary CRs B, Be, Li and $^{3}$He, when introducing nuisance scaling factors to renormalise the overall cross sections of production of B, Be and Li. The scaling factors obtained are of $\sim 1.18$, $\sim0.94$ and $\sim0.93$ for B, Be and Li, respectively, which are below the typical average experimental uncertainties of cross sections measurements for the best known channels.

Finally, from the study of the diffuse gamma-ray emission we have determined the need of including a break in the injection spectrum of electrons at a few GeV, inferred from the local HI gamma-ray emissivity and the diffuse IC emission at MeV energies.

%In conclusion, {\tt FLUKA} represents a promising tool for future more precise CR and multi-wavelength studies.

\section*{Acknowledgments}

P. De la Torre is supported by the Swedish National Space Agency under contract 117/19.
We acknowledge the {\tt FLUKA} collaboration for providing and supporting the code.
%This work has been partly carried out using the RECAS computing infrastructure in Bari (\url{https://www.recas-bari.it/index.php/en/}). A particular acknowledgment goes to G. Donvito and A. Italiano for their valuable support. This project used computing resources from the Swedish National Infrastructure for Computing (SNIC) under project Nos. 2021/3-42 and 2021/6-326 partially funded by the Swedish Research Council through grant no. 2018-05973.

%\section*{Appendix}

\section*{References}
\bibliography{biblio}

\begin{thebibliography}{10}

\bibitem{Luque:2021joz}
Pedro De~La Torre~Luque et~al.
\newblock {Implications of current nuclear cross sections on secondary cosmic
  rays with the upcoming DRAGON2 code}.
\newblock {\em JCAP}, 03:099, 2021.

\bibitem{Weinrich_combined}
N.~{Weinrich} and {others}.
\newblock {Combined analysis of AMS-02 (Li,Be,B)/C, N/O, $^{3}$He, and $^{4}$He
  data}.
\newblock {\em The Astrophysical Journal}, 639:A131, July 2020.

\bibitem{Korsmeier:2021brc}
M.~Korsmeier and A.~Cuoco.
\newblock {Implications of Lithium to Oxygen AMS-02 spectra on our
  understanding of cosmic-ray diffusion}.
\newblock {\em Phys. Rev. D}, 103(10):103016, 2021.

\bibitem{FLUKA}
Giuseppe~Battistoni et~al.
\newblock Overview of the fluka code.
\newblock {\em Annals of Nuclear Energy}, 82:10--18, 2015.
\newblock SNA + MC Conference, 2013.

\bibitem{Mazziotta:2020uey}
M.~N. Mazziotta and otherz.
\newblock {Cosmic-ray interactions with the Sun using the FLUKA code}.
\newblock {\em Phys. Rev. D}, 101(8):083011, 2020.

\bibitem{Mazziotta:2015uba}
M.~N. Mazziotta et~al.
\newblock {Production of secondary particles and nuclei in cosmic rays
  collisions with the interstellar gas using the FLUKA code}.
\newblock {\em AstroPart. Phys.}, 81:21--38, 2016.

\bibitem{Fermi-LAT:2016tkg}
M.~Ackermann et~al.
\newblock {Measurement of the high-energy gamma-ray emission from the Moon with
  the Fermi Large Area Telescope}.
\newblock {\em Phys. Rev. D}, 93(8):082001, 2016.

\bibitem{Luque:2022aio}
Pedro de la~Torre Luque et~al.
\newblock {FLUKA cross sections for cosmic-ray interactions with the DRAGON2
  code}.
\newblock 2 2022.

\bibitem{DRAGON2-1}
Carmelo Evoli et~al.
\newblock {Cosmic-ray propagation with $\small{DRAGON2}$: I. numerical solver
  and astrophysical ingredients}.
\newblock {\em JCAP}, 02:015, 2017.

\bibitem{DRAGON2-2}
Carmelo Evoli et~al.
\newblock {Cosmic-ray propagation with DRAGON2: II. Nuclear interactions with
  the interstellar gas}.
\newblock {\em JCAP}, 07:006, 2018.

\bibitem{Luque_MCMC}
P.~De La~Torre Luque et~al.
\newblock Markov chain monte carlo analyses of the flux ratios of b, be and li
  with the {DRAGON}2 code.
\newblock {\em JCAP}, 2021(07):010, jul 2021.

\bibitem{AMS02_BBeLi}
M.~Aguilar et~al.
\newblock Observation of new properties of secondary cosmic rays lithium,
  beryllium, and boron by the alpha magnetic spectrometer on the international
  space station.
\newblock {\em Phys. Rev. Lett.}, 120:021101, Jan 2018.

\bibitem{GALPROPXS}
Igor~V. {Moskalenko} and S.~G. {Mashnik}.
\newblock {Evaluation of Production Cross Sections of Li, Be, B in CR}.
\newblock In {\em ICRC}, volume~4 of {\em ICRC}, page 1969, July 2003.

\bibitem{webber2003updated}
WR~Webber, A~Soutoul, JC~Kish, and JM~Rockstroh.
\newblock Updated formula for calculating partial cross sections for nuclear
  reactions of nuclei with z$\leq 28$ and e$> 150$ mev nucleon--1 in hydrogen
  targets.
\newblock {\em ApJ Supplement Series}, 144(1):153, 2003.

\bibitem{silberberg1998updated}
R~Silberberg, CH~Tsao, and AF~Barghouty.
\newblock Updated partial cross sections of proton-nucleus reactions.
\newblock {\em ApJ}, 501(2):911, 1998.

\bibitem{FF}
L.~J. {Gleeson} and I.~H. {Urch}.
\newblock {A Study of the Force-Field Equation for the Propagation of Galactic
  Cosmic Rays}.
\newblock {\em Astrophysics and Space Science}, 25(2):387--404, December 1973.

\bibitem{Casandjian:2015hja}
Jean-Marc Casandjian.
\newblock {Local HI emissivity measured with Fermi-LAT and implications for
  cosmic-ray spectra}.
\newblock {\em Astrophys. J.}, 806(2):240, 2015.

\bibitem{Modified_FF}
Ilias Cholis, Dan Hooper, and Tim Linden.
\newblock A predictive analytic model for the solar modulation of cosmic rays.
\newblock {\em Phys. Rev. D}, 93:043016, Feb 2016.

\bibitem{Ghelfi}
A.~Ghelfi et~al.
\newblock {Neutron monitors and muon detectors for solar modulation studies: 2.
  \ensuremath{\phi} time series}.
\newblock {\em Adv. Space Res.}, 60(4):833--847, 2017.

\bibitem{Bernardo_2013}
Giuseppe~Di Bernardo et~al.
\newblock Cosmic ray electrons, positrons and the synchrotron emission of the
  galaxy: consistent analysis and implications.
\newblock {\em JCAP}, 2013(03):036--036, mar 2013.

\bibitem{CGRO}
{Siegert, T.}, {Berteaud, J.}, {Calore, F.}, {Serpico, P. D.}, and {Weinberger,
  C.}
\newblock Diffuse galactic emission spectrum between 0.5 and 8.0 mev.
\newblock {\em A\&A}, 660:A130, 2022.

\end{thebibliography}
\end{document}